\documentclass[11pt,a4paper]{article}
\usepackage[hyperref]{emnlp-ijcnlp-2019}
\usepackage{times}
\usepackage{latexsym}
\usepackage{comment}
\usepackage[colorinlistoftodos]{todonotes}

\usepackage{url}

\usepackage{graphics}      % for EPS, load graphicx instead 
\usepackage{subcaption}
\usepackage{verbatim}
\usepackage{color}
\usepackage{multirow}

\aclfinalcopy % Uncomment this line for the final submission

\title{Evaluation of Siamese Networks for Semantic Code Search}

\author{Raunak Sinha \\
  IBM Research \\
  \And
  Utkarsh Desai \\
  IBM Research \\
\And
  Srikanth Tamilselvam \\
  IBM Research \\
  \And
  Senthil Mani \thanks{ Work performed as part of IBM Research} \\
  Flipkart }
\date{}

\begin{document}
\maketitle
\begin{abstract}
With the increase in the number of open repositories and discussion forums, the use of natural language for semantic code search has become increasingly common. The accuracy of the results returned by such systems, however, can be low due to 1) limited shared vocabulary between code and user query and 2) inadequate semantic understanding of user query and its relation to code syntax. Siamese networks are well suited to learning such joint relations between data, but have not been explored in the context of code search. In this work, we evaluate Siamese networks for this task by exploring multiple extraction network architectures. These networks independently process code and text descriptions before passing them to a Siamese network to learn embeddings in a common space. We experiment on two different datasets and discover that Siamese networks can act as strong regularizers on networks that extract rich information from code and text, which in turn helps achieve impressive performance on code search beating previous baselines on $2$ programming languages. We also analyze the embedding space of these networks and provide directions to fully leverage the power of Siamese networks for semantic code search.
\end{abstract}

\section{Introduction}
Searching for code fragments is a very common activity in software development. The advent of large code repositories like GitHub\footnote{https://github.com/} and StackOverflow\footnote{https://stackoverflow.com/} has only increased the number of developers to rely on these repositories to search and reuse existing code \cite{reiss2009semantics}. Traditional Information Retrieval techniques \cite{lu2015query,lv2015codehow} do not work well for code search and retrieval tasks due to limited shared vocabulary between the source code and the natural language search text \cite{NCS}. Often, developers who are new to a programming language, search for code snippets in a context-free natural language. The choice of words used to search may not overlap with the code snippets leading to failure of traditional information retrieval systems. Therefore, there is a need to gain a deeper understanding of code and text in order to find semantically relevant code snippet.

Consider an example where a developer has a functional requirement to validate if \textit{age} is always lesser than $99$ and alert otherwise. The developer is tasked to enforce this check in Java. A naive Java developer who is not familiar with the language might make a query based on the requirement as: \textit{java check condition correctness}. The top 10 results\footnote{\label{footnote:date}As of December 9, 2019} in StackOverflow do not discuss the \textit{assert} keyword. A more programming friendly query such as \textit{java boolean check} or the \textit{assert} keyword itself results in code snippets demonstrating the steps as the top result in StackOverflow.

Use of deep neural network models have shown tremendous improvements in many tasks across domains including language tasks \cite{luong2015stanford,mishra2018cognition,gururangan2019variational}. This success can be largely attributed, in part, to their ability to learn meaningful relationships among words in documents efficiently and represent them in a way such that semantically equivalent words tend to have similar representations \cite{mikolov2013efficient, mikolov2013distributed}. One such family of models that are popular for determining text similarity are Siamese networks. First introduced by \cite{bromley1994signature}, a typical Siamese network consists of two identical sub networks that share weights. They work in tandem on different inputs and the output of both the networks are evaluated by a distance measure that also acts as a scoring function. This has been successfully applied in many similarity tasks in image domain \cite{taigman2014deepface, koch2015siamese, schroff2015facenet} and recently in text domain as well \cite{mueller2016siamese, neculoiu2016learning, das-etal-2016-together}. Another useful property of these models is their capability to learn from fewer data examples \cite{koch2015siamese}. Since code can be treated as a special kind of text data, one possible way to approach the problem of Semantic Code Search (commonly also referred to as \textit{Code Retrieval}) is to treat it as a similarity task where the objective is to bring semantically equivalent code snippets and their natural language descriptions closer. Therefore, we study the application of Siamese networks to code and corresponding text descriptions for semantic code search.

We apply multiple variations of the base Siamese network model on two different datasets for semantic code search and study its efficacy.  We further take the state of the art baselines - \cite{DCS} and \cite{CoaCor} on these datasets and observe that Siamese networks can improve over the baseline results invariably (upto 16\% improvement in MRR on one of the datasets). Finally, we present our analysis on the  performance of different Siamese network architectures explored and identify the conditions for improved performance.

The rest of the paper is organized as follows. We introduce some relevant prior art in section $2$. Next, in section $3$, we provide some background on Siamese networks and semantic code search and introduce terminology. In section $4$, we describe our approach and the different architectures investigated. In section $5$, we describe our experiments and present the results. Finally in section $6$, we perform a detailed analysis of our observations, followed by conclusions in section $7$.

% \tikz \draw[] (0,0) rectangle (\linewidth, -1in) node[pos=.2]{Answer Here:};

\section{Related Work}

Traditionally solutions to code search were based on information retrieval techniques and natural language processing comprising of query expansion and reformulation \cite{lu2015query,lv2015codehow}. \cite{lu2015query} expanded the query with synonyms from wordnet to search for code snippets.  API documentation was leveraged for query expansion for code snippets retrieval by the CodeHow tool proposed by \cite{lv2015codehow}.  The fundamental drawback of the above techniques is that there is a disparity in word overlap, including their synonyms, between the intent expressed in natural language and the low-level implementation details in the source code. There is a need to semantically relate the words between the two domains.
% Traditionally code search is based on information retrieval techniques. \cite{lu2015query} expanded the query with synonyms from wordnet to search for code snippets. \cite{lv2015codehow} expanded the query with relevant API documentation and further applied the extended boolean model to retrieve the code snippets.

Lately deep learning techniques have been administered for understanding code semantics and structure \cite{CodeNN,DCS,NCS,Code2Seq,CoaCor,UNIF,Code2Vec}. approaches the task of code summarization through an attention-based recurrent framework. The model generates a summary for the given code snippet which is then used to rank the relevance of the code to queries. 

Deep Code Search (DCS) \cite{DCS} follows a slightly different approach. DCS takes three aspects of code namely the a) method name, b) API invocation sequence, and c) the tokens and in parallel also takes the code descriptions as inputs to a different network and learns corresponding embeddings. Similarity between the embeddings are measured using cosine similarity. However, learning between embeddings is not explicitly shared. In one of our experiments we apply a Siamese network on top of DCS to combine the embedding learning framework with a Siamese style of sharing parameters between the two sub networks.

Neural Code Search (NCS) \cite{NCS} is an unsupervised model that proposed a way to aggregate vector representation of the source code using TF-IDF weighting to form document embeddings. It uses FastText \cite{fast-text} for learning the embeddings for these bags of words.  Further, the similarity between these embeddings are used for retrieval. Embedding Unification (UNIF) \cite{UNIF} is an extension over NCS, applying attention on the code tokens. Again, both these works treat code and text as independent learning modules and project them in a common high dimension space. 

CoaCor \cite{CoaCor} proposed an Reinforcement Learning framework to generate code annotations and show the improvements on code retrieval (CR) when combined with existing CR models like DCS \cite{DCS}. Though the objective is similar, we study Siamese for joint learning on top of DCS without generation.

Siamese Networks are popular for tasks that involve finding similarity or a relationship between two comparable artifacts. Introduced first in images \cite{bromley1994signature}, it has been applied in text domain to score relevance between a question and an answer candidate \cite{yin2016abcnn,das-etal-2016-together} and for learning text similarity \cite{mueller2016siamese, neculoiu2016learning}. But to the best of our knowledge, we are not aware of any work on applying Siamese networks for semantic code search.
\section{Background}

\begin{figure}[!t]
    \centering
	\includegraphics[width=.4\textwidth]{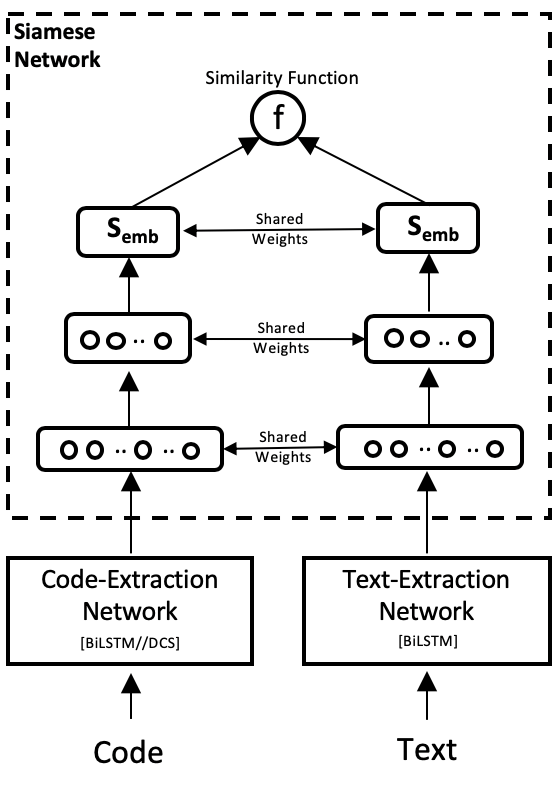}
	\caption{Architecture for Code Retrieval.}
	\label{fig:architecture}
\end{figure}

We now introduce some background relevant to our work in this paper.
% \subsection{Code Retrieval}
% Although Code Retrieval has been studied using Information Retrieval approaches [ref1][ref2], we will focus on Neural Network based methods. These methods typically aim to map both code snippets and description texts into a common embedding space such a given  code snippet and its description have similar vectors - with similarity defined using one of the methods listed in the next section. We consider two prominent Deep learning models in this work. 
% The model of [dcsref] (we will refer to it as DCS model), uses separate networks to encode a function and its corresponding description text. The text encoder is a simple RNN followed by a max-pooling layer to produce a text embedding. The code encoder consists of separate networks to encode method name, API sequences and tokens in code which are then combined into a final embedding. The DCS model tries to maximize the cosine similarity between these two embeddings.
% The network proposed in [codenn ref], called CODE-NN, 
 \subsection{Siamese Networks}
 As mentioned abovSiamese networks have been successfully applied to a variety of tasks that require jointly learning embeddings \cite{mueller2016siamese, neculoiu2016learning, yin2016abcnn,das-etal-2016-together}. The primary idea behind Siamese networks is of two twin or identical networks that share weights and are connected at the top via an objective function. A pair of inputs is one to each twin network is expected to learn a similar representation for inputs. Several variations of this setup have been developed for various input modalities \cite{koch2015siamese, mueller2016siamese, abdelpakey2019domainsiam} and similarity functions \cite{koch2015siamese, hoffer2015deep, das-etal-2016-together}.
 
 This property of jointly learning an embedding for a pair of inputs is particularly relevant for semantic code search which we explore in this work.
\subsection{Code Retrieval and treating Code as Text}
The objective of Semantic Code Search (henceforth referred to as simply, Code Retrieval) systems is to retrieve the most relevant code snippet(s) from a code repository in response to a natural language query or question by a user. A common way to approach the problem is to map all snippets in the repository to an embedding space. When a user query arrives, it is mapped to the same embedding space and the code snippet(s) closest to the query's embedding in this space are returned as the result. Learning an appropriate embedding is critical for this task.
As introduced in the previous section, a variety of methods have been explored for Code retrieval. Although Abstract Syntax Tree (AST) representations have also been used, a majority of these approaches process code as text for input to a deep neural network. 

Preprocessing is an imperative step for this task and several methods have been studied for representing code for input to deep neural networks. The most obvious way is to remove all code punctuation and tokenize code on whitespace. Language specific keywords are considered stop-words and are removed. We tokenize variable names and method names on camel-case and snake-case as these names may contain useful information about what the identifier is responsible for. Preprocessing done by \cite{DCS} looks at multiple representations of code for input to a deep network, such representations have been leveraged in multiple prior works and in our approach.

We now introduce some terminology. We refer to the code snippets from any code repository that are used during training as well as a response to the users' queries as simply \textit{code} or \textit{code snippet}. The code snippets can be a few lines of coherent code or entire function definitions. This distinction is irrelevant unless the function name is required by the model. We use the same term regardless. The natural language text that describes a code snippet (available in the documentation) or is the query to which the code snippet is an answer (mined from online forums) is referred to in this work as just \textit{text} or \textit{text description}. Thus, a [\textit{code,text}] pair can refer to any snippet of code and a natural language text that is  closely related to that code snippet.
\section{Approach}

In this work, we explore Siamese architectures for code retrieval where a code snippet is input to one branch of a Siamese network and the text description is input to the other. This distinction is important, in spite of the shared weights between the networks because even though we treat code as text, the code tokens can convey different meanings than text tokens. Therefore, all architectures we explore follow a common blueprint as shown in Figure \ref{fig:architecture}.

The code and text inputs are subjected to separate networks that process their respective input modalities in distinct ways (called extraction networks in the figure) and do not share weights. The outputs of these individual networks are passed to the Siamese network and a loss function computes the similarity between the representations learned by the two branches of the Siamese network. As highlighted in \cite{das-etal-2016-together}, this helps to capture the different expressions of the same concept in code and text. For example, the code snippet corresponding to the query \textit{``sort an array"} may not contain these terms or their synonyms. Thus, any candidate solution for code retrieval should have the capability to relate language semantics and code syntax.

With this motivation, we evaluate Siamese networks for code retrieval using $4$ pairs of architectures for the extraction networks as listed below. Each architecture pair is followed by a Siamese network that outputs an embedding of size $S_{emb}$. The code and text embeddings are compared using a loss function that encourages learning of similar embeddings for matching code and text pairs and diverse embeddings for unrelated code and text pairs. The candidates for such loss functions are discussed in section 4.2.

\begin{table*}[t]
\centering
\begin{tabular}{|c|c|c|c|c|c|c|}
\hline
\textbf{Dataset} & \textbf{Baseline} & \textbf{BiL-M} & \textbf{BiL-A} & \textbf{BiL-CS} & \textbf{DCS-Eval} & \textbf{Siamese-Eval} \\
\hline
(1) & (2) & (3) & (4) & (5) & (6) & (7) \\
\hline
Java 16M & 0.6 (DCS-Gu Model) & 0.294 & 0.180 & 0.318  &  \textbf{0.703} & 0.638*\\
StaQC SQL & 0.576 (CoaCor Model) & -- & 0.02 & 0.02 &  \textbf{0.593}& 0.208  \\
\hline
\end{tabular}
\caption{Baselines considered in this work for each of the datasets and the MRR on the test sets for the extraction networks followed by a Siamese network. Method names are not available for SQL and hence BiL-M model is not applicable. For the DCS-Siamese network, we perform evaluation on both the output of the DCS extraction network and the top Siamese layer. We report only the best performance over the different values of $S_{emb}$. The * indicates the reported number was obtained with $S_{emb}=200$ as opposed to $S_{emb}=2$ everywhere else. }
\label{tab:basicresults}
\end{table*}

\subsection{Model Architectures}
The base model for all our extraction networks for both code and text inputs is a BiLSTM (\textit{BiL}) network. The text extraction network takes the text as a sequence of tokens. However, the code extraction network can be provided with method names (\textit{M}), API calls (\textit{A}), or entire code snippet (\textit{CS}) as token sequences. 

In the first architecture \textbf{BiLSTMs-MethodName (BiL-M)}, we extract only the method name and tokenize it on camel-case and snake-case. The final hidden state vectors from code and text networks are provided as input to a Siamese network. This architecture cannot be applied if the code snippet is missing a method name. Instead of method name sequence, we provide the sequence of API calls in the code snippet as input to the code extraction networking after tokenization on camel-case and snake-case instantiating our second architecture \textbf{BiLSTMs-APIs (BiL-A)}. The third architecture \textbf{BiLSTMs-CodeSnippet (BiL-CS)} applies the same camel-case and snake-case tokenization on the entire code snippet.

%The text extraction network takes the text as a sequence of tokens. Alternatively, a sequence of code tokens (\textit{CT}) can be generated by tokenizing the entire code snippet instead of focusing on any code feature specifically. This gives us $3$ candidate architectures for the extraction networks.
% The final extraction network pair we explore treats code and text in completely different ways as described in \cite{gu2018deep}. 
 
%Our base architectures use only BiLSTMs for both code and text extraction networks. The text description is tokenized and presented to the text extraction network. For the code snippets, we extract only the method name and tokenize it on camel-case and underscores. This sequence of tokens is fed to the code extraction network \textbf{BiLSTMs-MethodName (BiL-M)}. The final hidden state vectors from these two networks are provided as input to a Siamese network. This architecture cannot be applied if the code snippet is missing a method name. 

%Instead of method name sequence, we provide the sequence of API calls in the code snippet as input to the code extraction networking after tokenization on camel-case and underscores instantiating our second architecture \textbf{BiLSTMs-APIs (BiL-A)}. The next architecture \textbf{BiLSTMs-CodeTokens (BiL-CT)} applies the same camel-case and underscore tokenization on the entire code snippet.

The next architecture - \textbf{Deep Code Search (DCS)} is based on the model presented in \cite{DCS}.  The code extraction network comprises of individual networks that process method names, API sequences and Bag-of-Words tokens and their representations are then combined using max pooling. The text extraction network is a simple BiLSTM. This setup has shown impressive results for the code retrieval task and their performance has been further surpassed by \cite{CoaCor} which also leverages the same setup. 

For evaluating any model given a correct [code,text] pair, we compute the text embedding by passing it through the network. We then compare the text embedding with all of the code embeddings and rank them based on cosine similarity. Note that the embeddings at any layer can be considered for similarity computation. For the first three extraction networks (BiL-M, BiL-A, BiL-CS), the final layer of the Siamese network is used for evaluation. However, for the DCS extraction network, we perform evaluation on both the output of the final Siamese layer and the output of the DCS extraction network. This gives the final two model architectures, \textbf{DCS-Eval} and \textbf{Siamese-Eval}, differing in what layer is used for evaluation/retrieval purposes. When this distinction is irrelevant, we refer to the model leveraging the DCS extraction network with a Siamese network as \textbf{DCS-Siamese}.

%We feel this simple setup does an excellent job at capturing relevant features from code without looking at detailed code structure in the form of Abstract Syntax Tree representations, for example.
% \end{itemize}

For the above $5$ models, we experimented with multiple sizes of $S_{emb}$ and the results are presented in the next section. %We present our experiments in the next section followed by a detailed discussion of our findings.

\subsection{Loss functions}
% explain what loss function we end up using and why. might also mention 1-2 lines about other possible loss functions in the context of siamese

We explore three optimization functions for jointly learning the embeddings. 

\begin{itemize}
    \item \textbf{Contrastive Loss}: This distance based loss performs pair-wise optimization of code snippets and text descriptions \cite{contrastive-loss}.
    
    \item \textbf{Triplet Loss}: Distance is calculated between an anchor (code-snippet), a positive candidate (related description) and a negative candidate (unrelated description). The distance between positive candidate and anchor is minimised, whereas, the distance between negative candidate and anchor is maximised \cite{ hoffer2015deep,triplet-loss}.
    
    \item \textbf{Cosine-Contrastive loss}: This is a trade-off between cosine and contrastive loss functions. Since the responses to a query are retrieved using cosine distance during inference, we use the cosine distance during training as well. Additionally, the contrastive component ensures pair-wise optimisation between code and descriptions.
\end{itemize}

    %\[ L = \lambda_1(cosine-eqn) + \lambda_2(contrastive-eqn)\]
    %We experiment with combinations weights and find our model to perfrom best for $\lambda_1$ = $\lambda_1$ = 0.5.
% \end{itemize}

\section{Experiments}

% \begin{table}
% \small
% \centering
% \begin{tabular}{|p{0.3\columnwidth}|p{0.2\columnwidth}|p{0.3\columnwidth}|}
% %\hline
% \textbf{Dataset} & \textbf{MRR}  & \textbf{Model}\\
%  &  \textbf{Baseline} & \\
% \hline
% Java-16M & 0.6 & DCS-Gu Model \\
% StaQC SQL & 0.576 & CoaCor Model\\
% \hline
% \end{tabular}
% \caption{Benchmark MRR and the methods that achieve them}
% \label{tab:baselines}
% \end{table}

We now describe the datasets used in our experiments, the training and evaluation processes in our setup, followed by our results.

\subsection{Datasets and Baselines}
One of the primary concerns while evaluating a code retrieval approach is the lack of standard datasets. Most approaches prefer collecting their own dataset and compare against some baseline that can be applied on the newly collected data, which again might not be openly available. This makes it extremely difficult to compare approaches on uniform grounds. Limited by this issue, we selected two datasets that have been evaluated on atleast two separate approaches. These datasets represent two different kinds of languages (strongly typed and declarative) and have enough variation to test the general applicability of our approach. 

The first dataset consists of $16M$ Java method definitions and corresponding descriptions crawled from Github \footnote[3]{https://github.com/guxd/deep-code-search}. \cite{DCS} reported impressive results on this dataset. We refer to their model as the \textit{DCS-Gu} model. We selected this dataset due to the large training set size. We also use a reduced version of this dataset comprising of 1\% of the data to explore and filter multiple architectures in our experiments. The limitation of this dataset is that it is only available in processed form and the original code-description pairs are not publicly available. 

The next dataset is the StaQC SQL dataset comprising of $119K$ SQL queries and question title pairs extracted from StackOverflow \footnote[4]{https://github.com/LittleYUYU/StackOverflow-Question-Code-Dataset}. The \textit{CoaCor} model \cite{CoaCor} achieves superior performance on this dataset surpassing the methods of \cite{DCS} and \cite{CodeNN}. Their approach also uses the DCS network of \cite{DCS} as a component. The baselines we use for each dataset are summarized in Table \ref{tab:basicresults}. Mean Retrieval Rank (MRR) is the commonly used metric for Code retrieval and is used for all our evaluations. 

% We evaluate our approach on three popular  model architectures for Code Retrieval. Each of these models learns a separate embedding for code and the corresponding text annotation, making them compatible inputs for the Siamese network. The DCS model (\textbf{todo}: reference) learns code embedding by extracting 3 sets of features from the code and passing them through separate networks to obtain individual embeddings which are then combined into a single embedding vector. The embedding for text is obtained via a Bi-LSTM network. The CoaCor model (\textbf{todo}:reference) uses Reinforcement learning to learn a Code Retrieval model using a Code Annotation model. We build the Siamese network over the QC-based code retrieval model in the CoaCor setup. (\textbf{todo}: add the 3rd model once decided).

% We report our results on the same datasets corresponding to the models described above and compare with the results from the base models. That is, we evaluate our approach on the Java dataset with 16M code-text pairs used to train the original DCS model and compare our results with the results reported in \textbf{dcs refrence}. Similarly, we use the SQL dataset introduced in \textbf{coacor reference} and compare our results with the CoaCor model. \textbf{todo: add last model/dataset}.

% Table X shows the benchmark performance metrics on each of these datasets and the models that have reported these numbers.

\begin{figure*}[t]
	\begin{center}
	\includegraphics[width=0.8\textwidth]{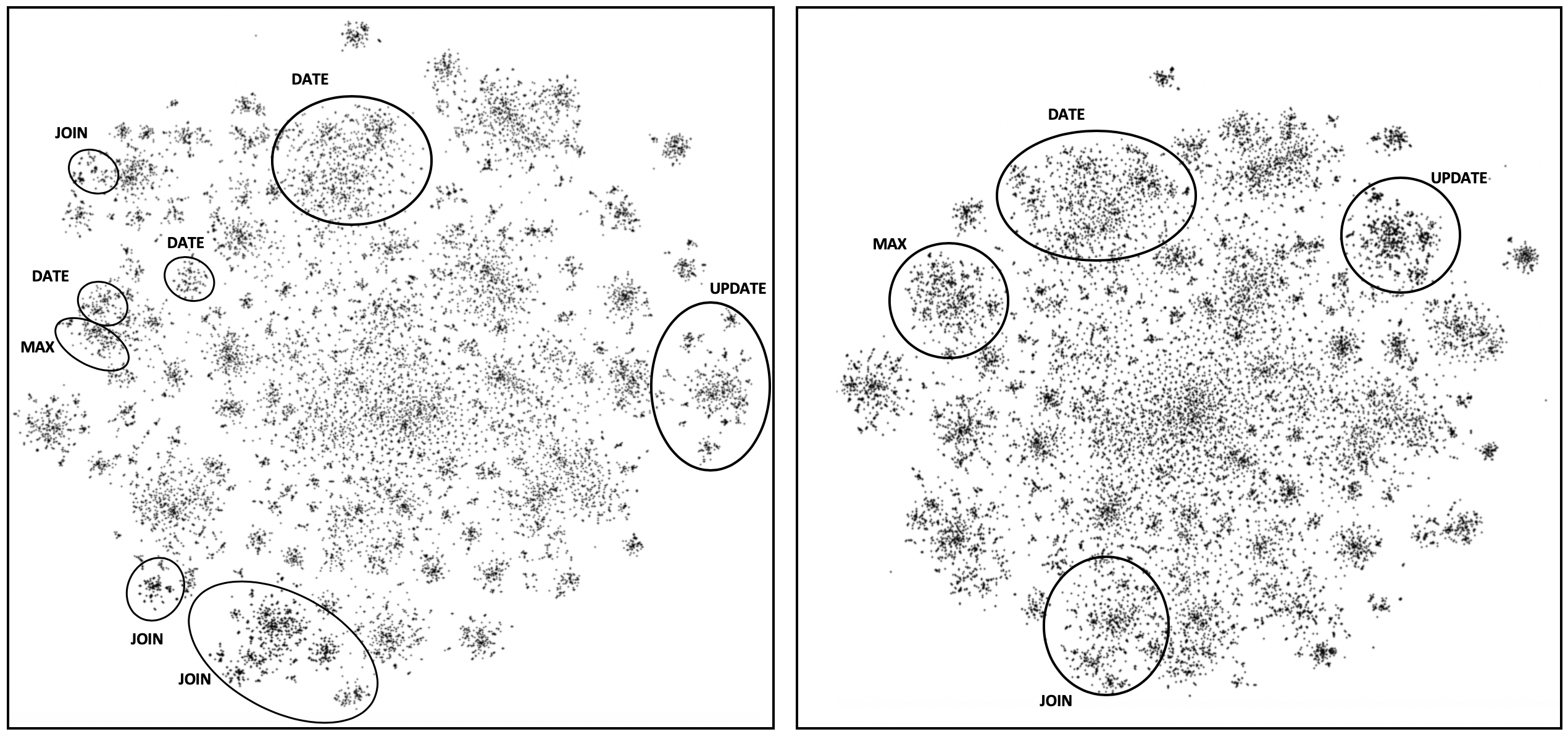}
	\end{center}
	\caption{t-SNE visualizations of the embeddings learnt by the DCS network (left) as described in \cite{DCS} compared to the embeddings produced by the DCS extraction network after using a Siamese network on top (right). The points in the figure represent texts in StaQC SQL dataset. Some clusters are highlighted.}
	\label{fig:SembComparison}
\end{figure*}

\subsection{Experimental Setup}
For all our experiments on the extraction networks, we experimented with multiple architectures for the Siamese network using the different loss functions. We observed that the \textit{contrastive-cosine} loss performed better than the other loss functions. We also experimented with $S_{emb} = {\{2, 100, 200\}}$ for each of  these architectures. For Siamese networks that have $S_{emb}=200$, we use layers of sizes $\{400, 300\}$ after the extraction network. For $S_{emb}=100$, an additional layer of size $200$ was used and for $S_{emb}=2$, another layer of size $50$ is used. When evaluating the DCS extraction network, the first layer of the Siamese network is of size $800$, to account for the output dimensionality of DCS. All layers are followed by ReLU and BatchNorm layers. 

The extraction network pair and its corresponding Siamese network are trained in an end-to-end fashion with Adam optimizer, initial learning rate $0.001$ and patience $40$. After $40$ epochs of stagnant performance, the learning rate is halved and the training resumed. This is repeated up to 4 times.

\subsection{Results}
Table \ref{tab:basicresults} (columns 3, 4 and 5) lists the MRR values of the best performing models for retrieval on the test sets for the first 3 extraction network architectures.
None of the architectures were able to surpass the baseline performance on the two datasets. With this experiment, for the given datasets, we find that Siamese networks by themselves were not able to achieve a comparable performance with any of the baselines. 

Further, we observed that the performance of these architecture was influenced mostly by the choice of $S_{emb}$ while the number and sizes of the intermediate layers did not have any notable effect. These models consistently achieved their best performance for $S_{emb}=2$.

%We report the MRR values of the best performing models following our setup and discuss them in detail in the next section.

% the models consistently achieved their best performance for $S_{emb}=2$. We analyze this behaviour in detail in the next section. Thus, Siamese networks by themselves have not been able to achieve a comparable performance to any of the baselines. However, as expected, there appears to be a direct correlation between the complexity of code features input to the model and it's performance.

Table \ref{tab:basicresults} (columns 6 and 7) summarizes the performance of the Siamese network with DCS extraction network on the two datasets. Not only is the retrieval MRR on the test sets observed to be better than any of the other architectures, the DCS-Eval model consistently outperforms the baseline models on each of the datasets. However, in this case, we observed that the best performing models were not always with $S_{emb}=2$. The Siamese-Eval model when trained on the Java $16M$ dataset and evaluated at the output of the Siamese layer performed better for $S_{emb}=200$. We denoted this with an *. 

% \begin{table}[th]
% \small
% \centering
% \begin{tabular}{|c|c|c|c|}
% \hline
% \textbf{Dataset} & \textbf{Baseline} & \textbf{Siamese-Eval} & \textbf{DCS-Eval} \\
% \hline
% Java 16M & 0.6 & 0.638* & \textbf{0.703} \\
% StaQC SQL & 0.576 & 0.208 & \textbf{0.593} \\
% \hline
% \end{tabular}
% \caption{Performance of the DCS-Siamese model on the 2 datasets when evaluated on the outputs of the DCS layer and the top Siamese layer. We report only the best performance over the different values of $S_{emb}$. The * indicates the reported number was obtained with $S_{emb}=200$ as opposed to $S_{emb}=2$ everywhere else.}
% \label{tab:ourresults}
% \end{table}

\begin{table}[]
\small
\centering
\begin{tabular}{|p{0.2\columnwidth}|p{0.7\columnwidth}|}
\hline
\textbf{Cluster Name} & \textbf{Sample Questions} \\
\hline
\multirow{2}{*}{MAX} & Get value based on max of a different column grouped by another column \\
 & How to query row with max value in a column in a referencing table ? \\
 \hline
\multirow{2}{*}{DATE} & How to use if and else to check the hour and run a specific query? \\
 & Counting the number of days excluding Sunday between two dates \\
 \hline
\multirow{2}{*}{JOIN} & How to prevent join from matching same row over and over again? \\
 & How do i pull multiple items out of a table with one join? \\
 \hline
\multirow{2}{*}{UPDATE} & How to update a row from another row of same table? \\
 & Mysql update attribute only when value is not null \\
 \hline
\end{tabular}
\caption{Some example questions from the clusters shown in t-SNE comparison in Figure \ref{fig:SembComparison}. }
\label{tab:clusternames}
\end{table}

\begin{figure*}[t]
\centering
  \includegraphics[width=0.8\textwidth]{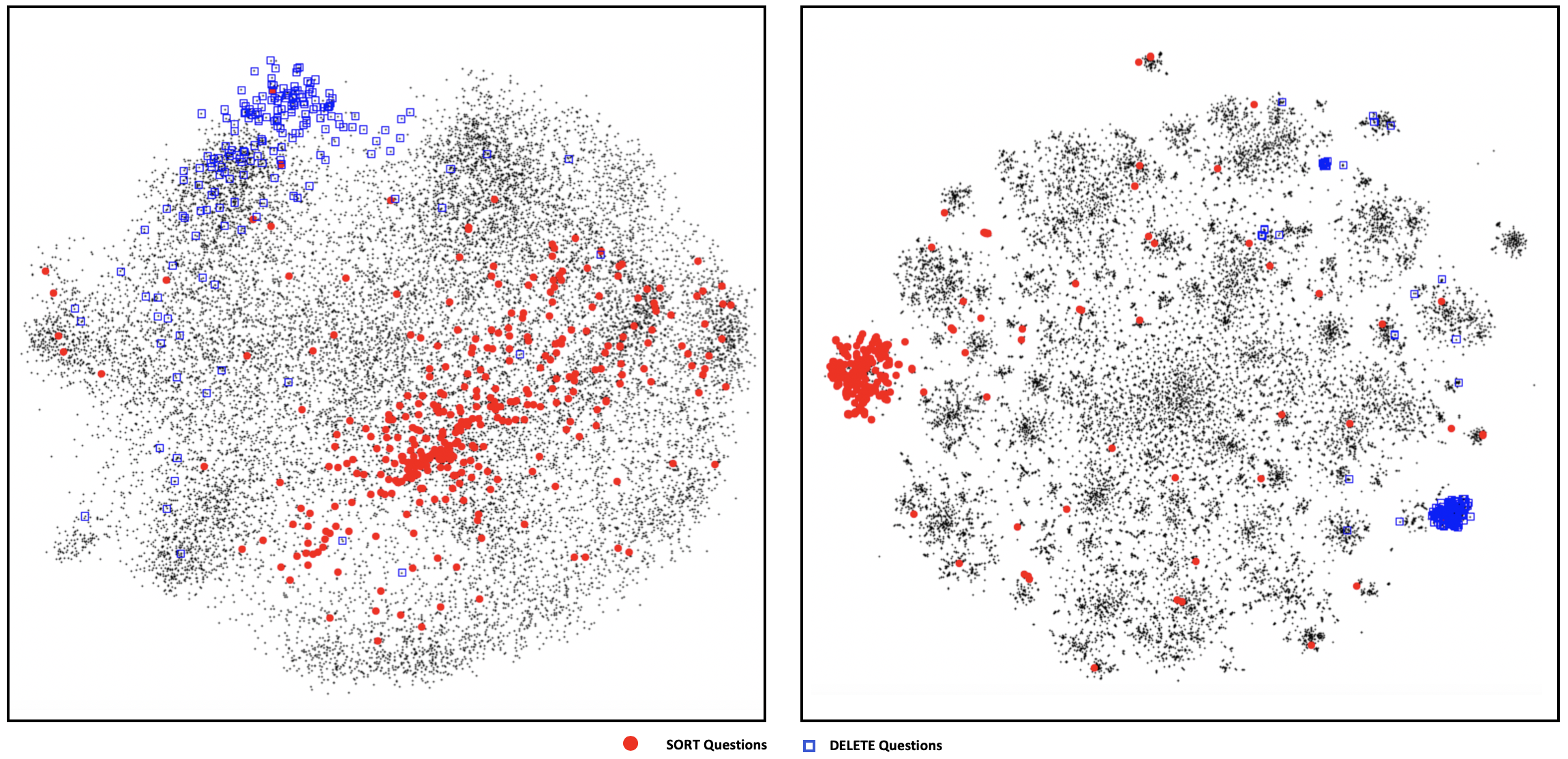}
\caption{t-SNE visualizations of the Siamese (left) and DCS layer embeddings (right) of the DCS-Siamese network with $S_{emb}=2$ for two sets of questions. The blue squares represent questions about deleting rows. The red circles are questions on sorting and ordering.}
\label{fig:emb_tsneplots_semb2}
\end{figure*}

\begin{figure*}[]
\centering
  \includegraphics[width=0.8\textwidth]{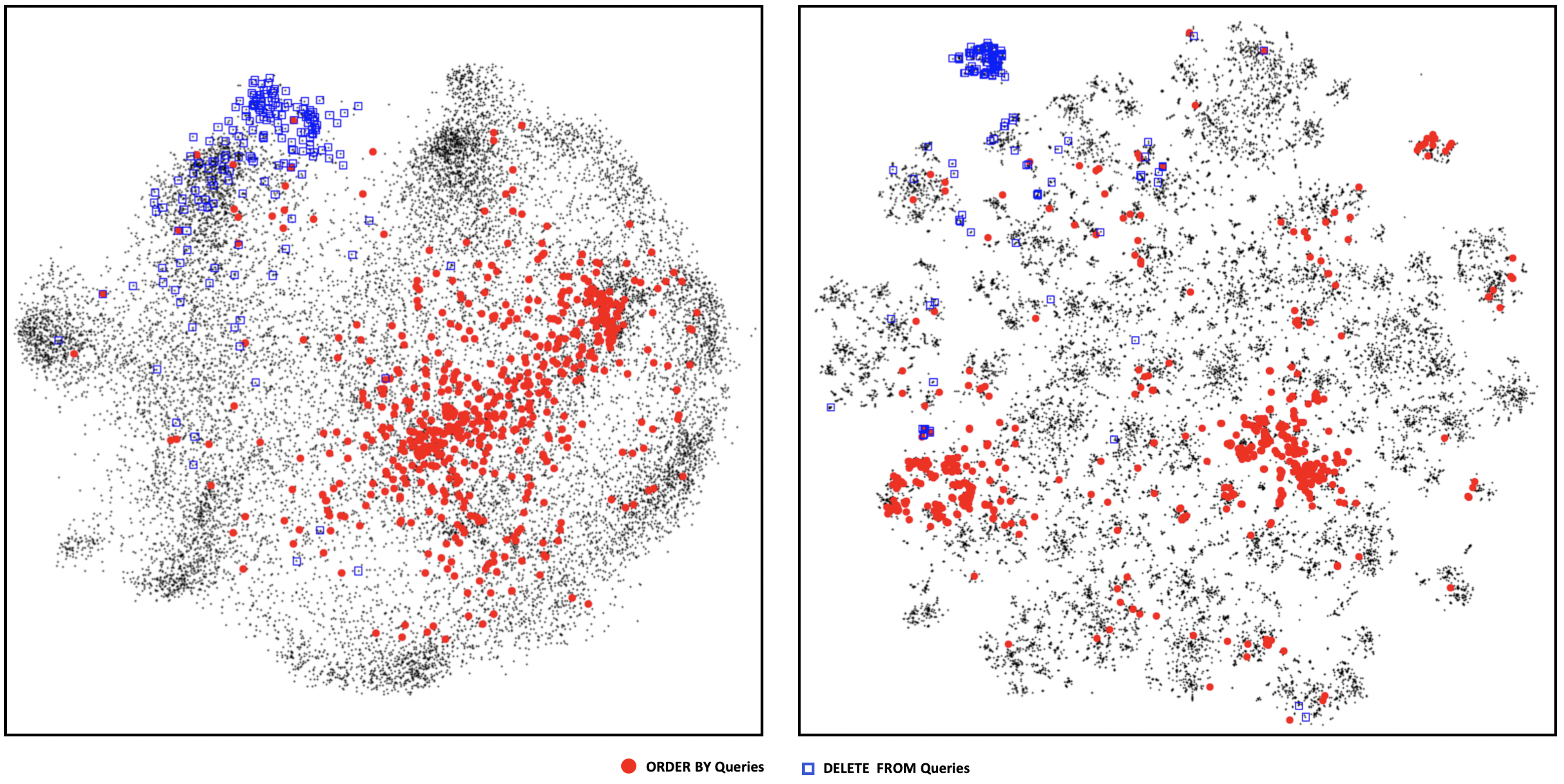}
 \caption{t-SNE visualizations of the Siamese (left) and DCS layer embeddings (right) of the DCS-Siamese network with $S_{emb}=2$ for two sets of SQL queries. The blue squares represent 'DELETE FROM' queries. The red circles are queries containing 'ORDER BY'.}
\label{fig:code_tsneplots_semb2}
\end{figure*}

We summarize these results as follows:

% 1. The initial processing stages of the DCS model capture rich information from code more than any of the other extraction networks considered.
% 2. The Siamese network architecture that progressively compresses information through narrowing layers seems to guide the learning towards a better representation space.
% 3. Even with this guidance, the embeddings learnt at a lower layer are more beneficial for the task as compared to the final layer of the network.
1. The DCS extraction network with a Siamese network performs better than the rest of the extraction networks considered and outperforms the baselines on the two datasets.

2. This superior performance is achieved when the retrieval is performed using embeddings at the output of the DCS extraction network, rather than at the output of the Siamese network.

3. Siamese network seem to have a regularizing effect that forces the extraction networks to learn good embeddings.

% \begin{figure*}[]
% \centering
% \begin{tabular}{cc}
% %   \includegraphics[width=40mm]{AuthorKit20/LaTeX/PLOTS/SST_combined_Bleu_CDF.png} & 
%   \includegraphics[width=.5\textwidth]{acl2020-templates/images/2siameselayer.png} & 
%  \includegraphics[width=.5\textwidth]{acl2020-templates/images/dcslayer.png}
% \end{tabular}
% \caption{t-SNE visualizations of the Siamese and DCS layer embeddings of the final network with Semb=2}
% \label{fig:emb_tsneplots_semb2}
% \end{figure*}

% \begin{figure*}[]
% \centering
% \begin{tabular}{cc}
% %   \includegraphics[width=40mm]{AuthorKit20/LaTeX/PLOTS/SST_combined_Bleu_CDF.png} & 
%   \includegraphics[width=.5\textwidth]{acl2020-templates/images/codeemb_sialayer.png} & 
%  \includegraphics[width=.5\textwidth]{acl2020-templates/images/codeemb_dcslayer.png}
% \end{tabular}
% \caption{t-SNE visualizations of the Code embeddings at the Siamese and DCS layer of the final network with Semb=2}
% \label{fig:code_tsneplots_semb2}
% \end{figure*}
\bibliography{emnlp-ijcnlp-2019}

\begin{thebibliography}{28}
\expandafter\ifx\csname natexlab\endcsname\relax\def\natexlab#1{#1}\fi

\bibitem[{Abdelpakey and Shehata(2019)}]{abdelpakey2019domainsiam}
Mohamed~H Abdelpakey and Mohamed~S Shehata. 2019.
\newblock Domainsiam: Domain-aware siamese network for visual object tracking.
\newblock In \emph{International Symposium on Visual Computing}, pages 45--58.
  Springer.

\bibitem[{Alon et~al.(2018)Alon, Brody, Levy, and Yahav}]{Code2Seq}
Uri Alon, Shaked Brody, Omer Levy, and Eran Yahav. 2018.
\newblock code2seq: Generating sequences from structured representations of
  code.
\newblock \emph{arXiv preprint arXiv:1808.01400}.

\bibitem[{Alon et~al.(2019)Alon, Zilberstein, Levy, and Yahav}]{Code2Vec}
Uri Alon, Meital Zilberstein, Omer Levy, and Eran Yahav. 2019.
\newblock code2vec: Learning distributed representations of code.
\newblock \emph{Proceedings of the ACM on Programming Languages}, 3(POPL):40.

\bibitem[{Bojanowski et~al.(2017)Bojanowski, Grave, Joulin, and
  Mikolov}]{fast-text}
Piotr Bojanowski, Edouard Grave, Armand Joulin, and Tomas Mikolov. 2017.
\newblock Enriching word vectors with subword information.
\newblock \emph{Transactions of the Association for Computational Linguistics},
  5:135--146.

\bibitem[{Bromley et~al.(1994)Bromley, Guyon, LeCun, S{\"a}ckinger, and
  Shah}]{bromley1994signature}
Jane Bromley, Isabelle Guyon, Yann LeCun, Eduard S{\"a}ckinger, and Roopak
  Shah. 1994.
\newblock Signature verification using a" siamese" time delay neural network.
\newblock In \emph{Advances in neural information processing systems}, pages
  737--744.

\bibitem[{Cambronero et~al.(2019)Cambronero, Li, Kim, Sen, and Chandra}]{UNIF}
Jose Cambronero, Hongyu Li, Seohyun Kim, Koushik Sen, and Satish Chandra. 2019.
\newblock When deep learning met code search.
\newblock \emph{arXiv preprint arXiv:1905.03813}.

\bibitem[{Das et~al.(2016)Das, Yenala, Chinnakotla, and
  Shrivastava}]{das-etal-2016-together}
Arpita Das, Harish Yenala, Manoj Chinnakotla, and Manish Shrivastava. 2016.
\newblock \href {https://doi.org/10.18653/v1/P16-1036} {Together we stand:
  {S}iamese networks for similar question retrieval}.
\newblock In \emph{Proceedings of the 54th Annual Meeting of the Association
  for Computational Linguistics (Volume 1: Long Papers)}, pages 378--387,
  Berlin, Germany. Association for Computational Linguistics.

\bibitem[{Gu et~al.(2018)Gu, Zhang, and Kim}]{DCS}
Xiaodong Gu, Hongyu Zhang, and Sunghun Kim. 2018.
\newblock Deep code search.
\newblock In \emph{2018 IEEE/ACM 40th International Conference on Software
  Engineering (ICSE)}, pages 933--944. IEEE.

\bibitem[{Gururangan et~al.(2019)Gururangan, Dang, Card, and
  Smith}]{gururangan2019variational}
Suchin Gururangan, Tam Dang, Dallas Card, and Noah~A Smith. 2019.
\newblock Variational pretraining for semi-supervised text classification.
\newblock \emph{arXiv preprint arXiv:1906.02242}.

\bibitem[{Hadsell et~al.(2006)Hadsell, Chopra, and LeCun}]{contrastive-loss}
Raia Hadsell, Sumit Chopra, and Yann LeCun. 2006.
\newblock Dimensionality reduction by learning an invariant mapping.
\newblock In \emph{2006 IEEE Computer Society Conference on Computer Vision and
  Pattern Recognition (CVPR'06)}, volume~2, pages 1735--1742. IEEE.

\bibitem[{Hoffer and Ailon(2015)}]{hoffer2015deep}
Elad Hoffer and Nir Ailon. 2015.
\newblock Deep metric learning using triplet network.
\newblock In \emph{International Workshop on Similarity-Based Pattern
  Recognition}, pages 84--92. Springer.

\bibitem[{Iyer et~al.(2016)Iyer, Konstas, Cheung, and Zettlemoyer}]{CodeNN}
Srinivasan Iyer, Ioannis Konstas, Alvin Cheung, and Luke Zettlemoyer. 2016.
\newblock Summarizing source code using a neural attention model.
\newblock In \emph{Proceedings of the 54th Annual Meeting of the Association
  for Computational Linguistics (Volume 1: Long Papers)}, pages 2073--2083.

\bibitem[{Koch et~al.(2015)Koch, Zemel, and Salakhutdinov}]{koch2015siamese}
Gregory Koch, Richard Zemel, and Ruslan Salakhutdinov. 2015.
\newblock Siamese neural networks for one-shot image recognition.
\newblock In \emph{ICML deep learning workshop}, volume~2.

\bibitem[{Lu et~al.(2015)Lu, Sun, Wang, Lo, and Duan}]{lu2015query}
Meili Lu, Xiaobing Sun, Shaowei Wang, David Lo, and Yucong Duan. 2015.
\newblock Query expansion via wordnet for effective code search.
\newblock In \emph{2015 IEEE 22nd International Conference on Software
  Analysis, Evolution, and Reengineering (SANER)}, pages 545--549. IEEE.

\bibitem[{Luong and Manning(2015)}]{luong2015stanford}
Minh-Thang Luong and Christopher~D Manning. 2015.
\newblock Stanford neural machine translation systems for spoken language
  domains.
\newblock In \emph{Proceedings of the International Workshop on Spoken Language
  Translation}, pages 76--79.

\bibitem[{Lv et~al.(2015)Lv, Zhang, Lou, Wang, Zhang, and Zhao}]{lv2015codehow}
Fei Lv, Hongyu Zhang, Jian-guang Lou, Shaowei Wang, Dongmei Zhang, and Jianjun
  Zhao. 2015.
\newblock Codehow: Effective code search based on api understanding and
  extended boolean model (e).
\newblock In \emph{2015 30th IEEE/ACM International Conference on Automated
  Software Engineering (ASE)}, pages 260--270. IEEE.

\bibitem[{Mikolov et~al.(2013{\natexlab{a}})Mikolov, Chen, Corrado, and
  Dean}]{mikolov2013efficient}
Tomas Mikolov, Kai Chen, Greg Corrado, and Jeffrey Dean. 2013{\natexlab{a}}.
\newblock Efficient estimation of word representations in vector space.
\newblock \emph{arXiv preprint arXiv:1301.3781}.

\bibitem[{Mikolov et~al.(2013{\natexlab{b}})Mikolov, Sutskever, Chen, Corrado,
  and Dean}]{mikolov2013distributed}
Tomas Mikolov, Ilya Sutskever, Kai Chen, Greg~S Corrado, and Jeff Dean.
  2013{\natexlab{b}}.
\newblock Distributed representations of words and phrases and their
  compositionality.
\newblock In \emph{Advances in neural information processing systems}, pages
  3111--3119.

\bibitem[{Mishra et~al.(2018)Mishra, Tamilselvam, Dasgupta, Nagar, and
  Dey}]{mishra2018cognition}
Abhijit Mishra, Srikanth Tamilselvam, Riddhiman Dasgupta, Seema Nagar, and
  Kuntal Dey. 2018.
\newblock Cognition-cognizant sentiment analysis with multitask subjectivity
  summarization based on annotators' gaze behavior.
\newblock In \emph{Thirty-Second AAAI Conference on Artificial Intelligence}.

\bibitem[{Mueller and Thyagarajan(2016)}]{mueller2016siamese}
Jonas Mueller and Aditya Thyagarajan. 2016.
\newblock Siamese recurrent architectures for learning sentence similarity.
\newblock In \emph{Thirtieth AAAI Conference on Artificial Intelligence}.

\bibitem[{Neculoiu et~al.(2016)Neculoiu, Versteegh, and
  Rotaru}]{neculoiu2016learning}
Paul Neculoiu, Maarten Versteegh, and Mihai Rotaru. 2016.
\newblock Learning text similarity with siamese recurrent networks.
\newblock In \emph{Proceedings of the 1st Workshop on Representation Learning
  for NLP}, pages 148--157.

\bibitem[{Reiss(2009)}]{reiss2009semantics}
Steven~P Reiss. 2009.
\newblock Semantics-based code search.
\newblock In \emph{Proceedings of the 31st International Conference on Software
  Engineering}, pages 243--253. IEEE Computer Society.

\bibitem[{Sachdev et~al.(2018)Sachdev, Li, Luan, Kim, Sen, and Chandra}]{NCS}
Saksham Sachdev, Hongyu Li, Sifei Luan, Seohyun Kim, Koushik Sen, and Satish
  Chandra. 2018.
\newblock Retrieval on source code: a neural code search.
\newblock In \emph{Proceedings of the 2nd ACM SIGPLAN International Workshop on
  Machine Learning and Programming Languages}, pages 31--41. ACM.

\bibitem[{Schroff et~al.(2015{\natexlab{a}})Schroff, Kalenichenko, and
  Philbin}]{schroff2015facenet}
Florian Schroff, Dmitry Kalenichenko, and James Philbin. 2015{\natexlab{a}}.
\newblock Facenet: A unified embedding for face recognition and clustering.
\newblock In \emph{Proceedings of the IEEE conference on computer vision and
  pattern recognition}, pages 815--823.

\bibitem[{Schroff et~al.(2015{\natexlab{b}})Schroff, Kalenichenko, and
  Philbin}]{triplet-loss}
Florian Schroff, Dmitry Kalenichenko, and James Philbin. 2015{\natexlab{b}}.
\newblock Facenet: A unified embedding for face recognition and clustering.
\newblock In \emph{Proceedings of the IEEE conference on computer vision and
  pattern recognition}, pages 815--823.

\bibitem[{Taigman et~al.(2014)Taigman, Yang, Ranzato, and
  Wolf}]{taigman2014deepface}
Yaniv Taigman, Ming Yang, Marc'Aurelio Ranzato, and Lior Wolf. 2014.
\newblock Deepface: Closing the gap to human-level performance in face
  verification.
\newblock In \emph{Proceedings of the IEEE conference on computer vision and
  pattern recognition}, pages 1701--1708.

\bibitem[{Yao et~al.(2019)Yao, Peddamail, and Sun}]{CoaCor}
Ziyu Yao, Jayavardhan~Reddy Peddamail, and Huan Sun. 2019.
\newblock Coacor: Code annotation for code retrieval with reinforcement
  learning.
\newblock In \emph{The World Wide Web Conference}, pages 2203--2214. ACM.

\bibitem[{Yin et~al.(2016)Yin, Sch{\"u}tze, Xiang, and Zhou}]{yin2016abcnn}
Wenpeng Yin, Hinrich Sch{\"u}tze, Bing Xiang, and Bowen Zhou. 2016.
\newblock Abcnn: Attention-based convolutional neural network for modeling
  sentence pairs.
\newblock \emph{Transactions of the Association for Computational Linguistics},
  4:259--272.

\end{thebibliography}
\bibliographystyle{acl_natbib}
% \input{appendix.tex}
%\

\end{document}